\documentclass[article,authoryear]{elsarticle}

\usepackage{lineno,hyperref}
\modulolinenumbers[5]

\journal{Journal of \LaTeX\ Templates}

\usepackage{graphics}
\usepackage{float,rotating}
\usepackage{wrapfig}
\usepackage{subcaption}

\usepackage{amsmath}
\usepackage{amsfonts}
\usepackage{amssymb}

\usepackage{mathrsfs}

\newcommand{\GeV}{\mbox{GeV}}
\newcommand{\TeV}{\mbox{TeV}}

\newcommand{\prof}{\mathcal{P}}
\newcommand{\ppterm}{\frac{d\overline{\Phi}^\textrm{PP}(E)}{dE}}
\newcommand{\sppterm}{d\overline{\Phi}^\textrm{PP}/dE}
\newcommand{\sv}{\langle \sigma \mathit{v} \rangle}
\newcommand{\aaa}{\alpha}
\newcommand{\tdm}{\tau}

\newcommand{\mdm}{m_\mathrm{DM}}

\newcommand{\q}{\mathscr{Q}}
\newcommand{\qac}{\mathscr{Q}_{Ac}}
\newcommand{\qle}{\mathscr{Q}_{Le}}
\newcommand{\qpsf}{\mathscr{Q}_{PSF}}
\newcommand{\qall}{\overline{\mathscr{Q}}}
\newcommand{\thetac}{\theta_{\text{c}}}
\newcommand{\Dom}{\Delta\Omega}
\newcommand{\DomON}{\Delta\Omega_{\text{ON}}}
\newcommand{\DomOFF}{\Delta\Omega_{\text{OFF}}}
\newcommand{\dom}{d\Omega}
\newcommand{\domON}{d\Omega_{\text{ON}}}
\newcommand{\domOFF}{d\Omega_{\text{OFF}}}

\begin{document}

\begin{frontmatter}

\title{Pointing optimization for IACTs on indirect dark matter searches}
% \tnotetext[mytitlenote]{Fully documented templates are available in the elsarticle package on \href{http://www.ctan.org/tex-archive/macros/latex/contrib/elsarticle}{CTAN}.}

%% Group authors per affiliation:
\author{J. Palacio\fnref{myemail}}
\fntext[myemail]{email to: jpalacio@ifae.es}
\author{D. Navarro-Girones}
\author{J. Rico}
\address{Institut de F\'isica d’Altes Energies (IFAE), The Barcelona Institute of Science and Technology
(BIST), E-08193 Bellaterra (Barcelona), Spain}

% more complex case: 4 authors, 3 institutions, 2 footnotes
% \author[a,1]{J. Palacio}
% \author[a]{D. Navarro-Girones,}
% \author[a]{and J. Rico}

% % The "\note" macro will give a warning: "Ignoring empty anchor..."
% % you can safely ignore it.
% \affiliation[a]{IFAE-BIST, Campus UAB, E-08193 Bellaterra, Spain}

% % e-mail addresses: one for each author, in the same order as the authors
% \emailAdd{jpalacio@ifae.es}

\begin{abstract}
We present a procedure to optimize the offset angle (usually also known as the \emph{wobble distance}) 
and the signal integration region for the observations and analysis of extended sources by Imaging Atmospheric
Cherenkov Telescopes (IACTs) such as MAGIC, HESS, VERITAS or (in the near future), CTA. 
Our method takes into account the off-axis instrument performance and the emission profile of the gamma-ray
source. 
We take as case of study indirect dark matter searches (where an a priori knowledge on the expected signal
morphology can be assumed) and provide optimal pointing strategies to perform searches of dark matter on 
a set of dwarf spheroidal galaxies with current and future IACTs.
\end{abstract}

\begin{keyword}
IACTs, off-axis performance, dark matter
\end{keyword}

\end{frontmatter}

%\linenumbers

\section{Introduction}
\label{sec:Introduction}
\noindent
Imaging Atmospheric Cherenkov Telescopes (IACTs) are ground based instruments 
capable of detecting gamma rays with energies 
from $\sim$50~\GeV ~to $\sim$100~\TeV.
IACT's typical fields of view (FoVs) are of the order of 
%the degree (
$\sim$1-10$^{\circ}$.
Observations are often performed in the so called
\emph{wobble mode}~\citep{Fomin:1994APh}, in which the nominal pointing of the telescope has an offset
(by a certain angle $w$, called the \emph{wobble distance})
w.r.t. the position of the source under observation
(or, for extended sources, to its center). 
\emph{Signal} (or ON) region is integrated inside a circular region of angular size $\theta_{\text{c}}$ around the source
while \emph{background control} (or OFF) region can be defined equally around a ghost region placed symmetrically
w.r.t. the pointing direction (in order to have equal acceptance).
%Under such method, \emph{signal} and
%\emph{background control} (or ON and OFF) regions of equal acceptance
%(because they are located symmetrically w.r.t the pointing direction) are observed simultaneously.
Under such wobble observation mode ON and OFF regions are observed simultaneously, 
what makes an efficient use of the limited duty cycles of IACTs while minimizing possible systematic differences
in the acceptance for ON and OFF regions (due e.g. to atmospheric changes in the on-axis observation mode). 
\newline

\noindent 
Unlike $\theta_{\text{c}}$ that is used in the analysis, 
$w$ is fixed during data taking (by fixing the pointing direction w.r.t. the center of the source).
The value of $w$ can be optimized if one takes into account that 
for large $w$, ON and OFF regions are defined close to the edge of the FoV, 
where the performance of the instrument decreases 
while for low $w$, it may not be possible to define an appropriate signal-free OFF region.  
These effects become critical for moderately extended sources, as the case for
instance of the expected gamma-ray signal coming from Dark Matter (DM) in nearby dwarf spheroidal galaxies (dSphs)
or from pulsar wind nebulae from nearby pulsars. 
\newline

\noindent
Here we present a procedure to optimize the wobble distance $w$ and signal integration radius $\theta_{\text{c}}$, 
taking into account the off-axis performance of the instrument and the expected spatial
morphology of the source.
% We should stress that, unlike $\theta_{c}$, which can be always modified during analysis,
%$w$ is fixed during data taking (by fixing the pointing direction w.r.t. the center of the source)
%and can not be modified a posteriori, and hence the importance of the method presented in here for planning observations. 
As a case study, we focus on indirect DM searches and provide optimal pointing configurations for a list of dSphs 
to be observed for current and future IACTs.
We have implemented an open-source tool %, freely distributed, 
so that the procedure can be applied to optimize
the pointing strategy of an arbitrary IACT observing an arbitrary 
circular symmetric moderate extended gamma-ray source.
\newline

\noindent
The rest of this paper is structured as follows: 
in \autoref{sec:JDInstrument} 
we introduce the IACT technique
and define a set of quantities that allow us to quantify their off-axis performance;
in \autoref{sec:JDOptimization} 
we introduce the quality factor that we use as a figure of merit for the optimization
of the pointing strategy;
in \autoref{sec:JDDarkMatter} we briefly discuss the DM paradigm and assess its framework,
and apply the method for the case of indirect DM searches to provide optimal pointing strategies  
on a set of dSphs observed with current or future IACTs;
finally, in \autoref{sec:Conclusions} 
we briefly discuss the current status of the software and its applicability.

\section{Imaging Atmospheric Cherenkov Telescopes off-axis performance}
\label{sec:JDInstrument}
\begin{figure}[t]
  \centering
  \includegraphics[width=.5\linewidth]{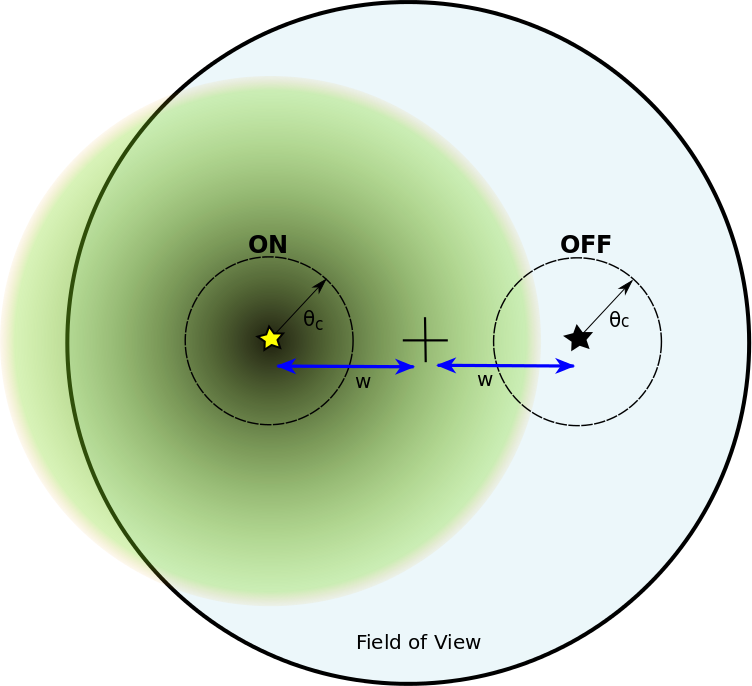}
  \caption[Wobble mode variables]{\label{fig:wobbleModeVariables} 
Schematic configuration of the FoV during \emph{wobble} mode observations.
The telescope pointing (black cross) has an offset distance \emph{w} w.r.t. the center of the 
source under study (yellow star). 
\emph{Signal} (\emph{ON}) region is defined as a circle around the center of the source, with angular size $\theta_{\text{c}}$.
One \emph{background control} region (circular region around OFF, black star) is defined with same angular size, 
symmetrically w.r.t. the signal region.
The leakage effect is schematically shown where, for moderately extended source (green area), 
signal events are also expected to be reconstructed inside OFF.}
\end{figure}
\noindent
In wobble operation mode, a circular ON region of radius $\theta_{\text{c}}$ is defined centered at the source under study
(observed at a distance $w$ from the center of the FoV, see~\autoref{fig:wobbleModeVariables}).
One or several OFF regions are defined within the same FoV, in such a way that background statistical uncertainties
are minimized and instrumental associated uncertainties are also kept low\footnote{
The response of the camera over the FoV is not perfectly homogeneous and different wobble strategies try to minimize
this effect.}.
For moderately extended sources, as the case we consider,
%, as the case we focus on in here, 
typically only a single OFF region is considered\footnote{
The study of the effect of different number of OFF regions is left as an improvement.} in order not to overlap ON and/or OFF regions
(circle ON and OFF in~\autoref{fig:wobbleModeVariables}).
\newline

\noindent
Due to their optics and trigger strategy, IACTs have a decreasing performance for detecting
gamma rays towards the edges of the FoV, w.r.t. its center (i.e. the pointing direction of the instrument).
In order to characterize the off-axis performance of IACTs,
we use the \emph{relative acceptance} ($\epsilon$)
w.r.t. the center of the FoV.
This relative acceptance can be estimated as;
\begin{eqnarray}\label{eq:relativeAcceptance}
\epsilon(d) = \frac{\mathcal{R_{\text{ON}}}(d)}{\mathcal{R_{\text{ON}}}(d=0)},
\end{eqnarray}
where $\mathcal{R}_{\text{ON}}$ is the rate of events passing all the analysis cuts 
(i.e. gamma-ray candidates) inside the ON region, and $d$ the off-set distance w.r.t. the pointing direction
(we assume $\epsilon$ to be circularly symmetric from the center of the FoV).
Note that, in~\autoref{eq:relativeAcceptance}, we are implicitly assuming $\theta_{\text{c}}$ to be much smaller than the scale of the FoV ($\theta_{\text{c}}\ll\sim5^{\circ}$),
otherwise, $\epsilon$ may vary from one point to another within the integration region.
As it will be used later on, 
we could have equally written~\autoref{eq:relativeAcceptance} replacing $\mathcal{R}_{\text{ON}}$ for $\mathcal{R}_{\text{OFF}}$,
with $\mathcal{R}_{\text{OFF}}$ being the rate of gamma-ray candidates inside the OFF region.
\subsection{Relative acceptance for real Imaging Atmospheric Cherenkov Telescope}
\label{subsec:AcceptanceMAGICandCTA}
\noindent
We compute now $\epsilon$ for %two real instruments operating now, or in a near future: 
the Florian Goebel Major Atmospheric Gamma-ray Imaging Cherenkov (MAGIC) telescopes and the future Cherenkov Telescope Array (CTA).
\newline

\begin{figure}[t]
\includegraphics[scale=0.4]{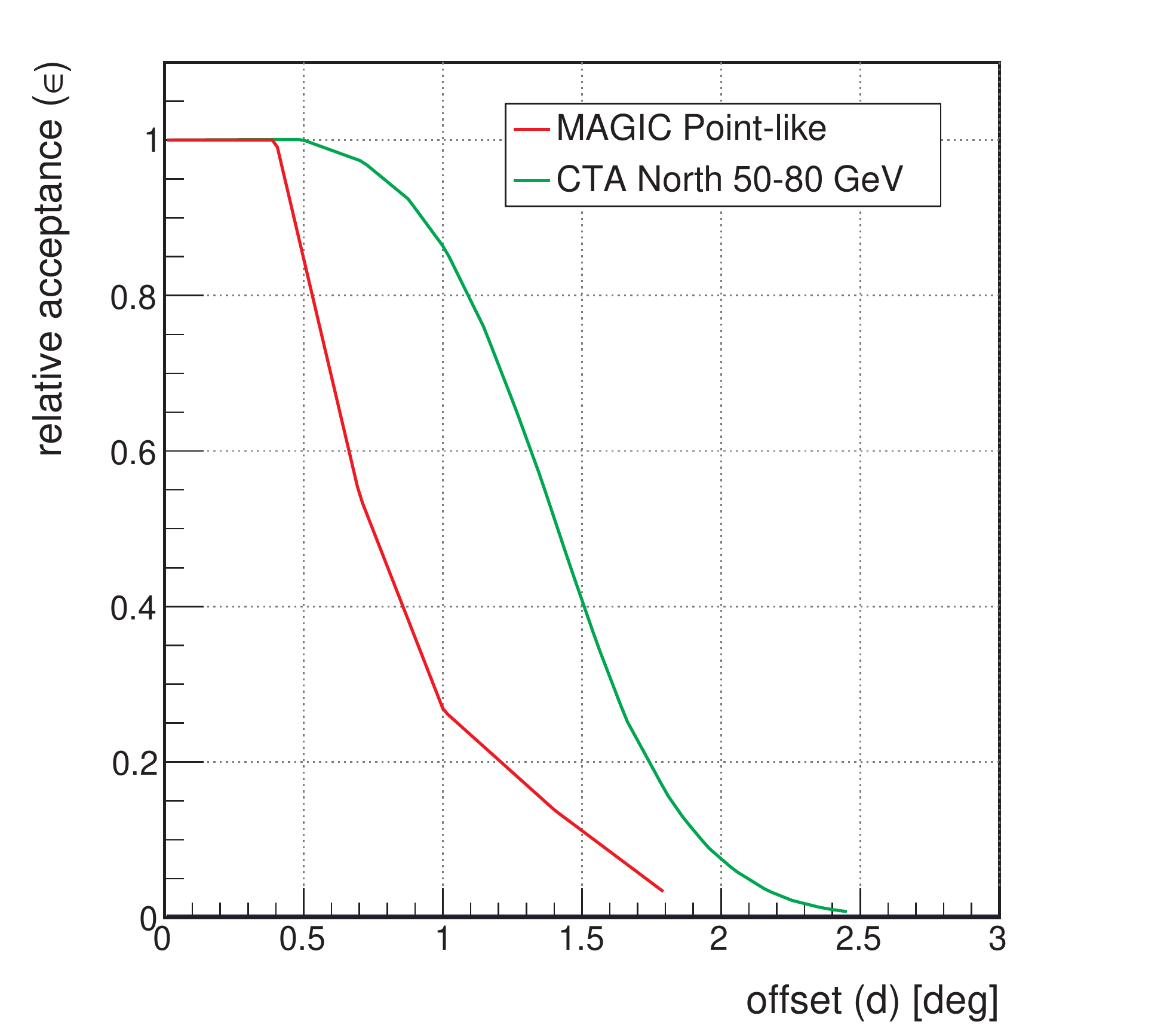}
\centering
\caption[MAGIC \& CTA relative acceptance]{\label{fig:relativeAcceptance}
$\epsilon_{\text{MAGIC}}$ (red) and $\epsilon_{\text{CTA}}$ (green) as a function of $d$ 
recomputed from~\citep{Aleksic:2014lkm} and \emph{https://www.cta-observatory.org/} respectively.}
\end{figure}

\noindent
MAGIC is a system of two gamma-ray Cherenkov telescopes located at the Roque de los Muchachos
Observatory in La Palma (Canary Islands, Spain), 
sensitive to gamma rays in the very high energy (VHE) domain, i.e.\ in the range between $\sim$50~$\GeV$ and $\sim$50~$\TeV$~\citep{Aleksic:2014lkm}.
The MAGIC FoV is $\sim3.5^\circ$ diameter.
Standard point-like observations are performed in wobble mode, with $(w,\theta_{\text{c}})_{\text{MAGIC}}=(0.4^\circ,\sim0.1^\circ)$.
Figure 20 in~\cite{Aleksic:2014lkm} shows the rate of gamma-like events detected from the direction of Crab-Nebula
observed at different values of $w$, for two different stable hardware configurations of MAGIC.
Using~\autoref{eq:relativeAcceptance}, we compute the relative acceptance of the MAGIC telescopes ($\epsilon_{\text{MAGIC}}$)
from the data from~\cite{Aleksic:2011bx} labeled as \emph{Crab Nebula post-upgrade}, hereafter named \emph{MAGIC Point-like} 
(see~\autoref{fig:relativeAcceptance}).
\newline

\noindent
CTA is the next generation ground-based observatory for gamma-ray astronomy at very-high
energies. %With more than 100 telescopes located in the northern and southern hemispheres,
CTA will be the world's largest and most sensitive high-energy gamma-ray observatory and will operate in both the northern
and southern hemispheres~\citep{Acharya:2017ttl}.
We take CTA's off-axis performance from 
\emph{https://www.cta-observatory.org/}, where the relative off-axis sensitivity
($\delta$) normalized to the center of the FoV is given. %for four different energy bins.
In order to compute the relative acceptance of CTA ($\epsilon_{\text{CTA}}$),
%\footnote{ 
%We are considering here $\epsilon$ of the full CTA array. In reality CTA will be formed by IACTs of different
%kind, each telescope type with a possible different $\epsilon$.
%}, 
we need to consider that $\delta$ can be written as
\begin{eqnarray}\label{eq:differentialSensitivity}
\nonumber    &\delta (d) =& \frac{\mathcal{S}(d)}{\mathcal{S}(d=0)} \; ,\\
\end{eqnarray}
where $\mathcal{S}$ is the sensitivity of the instrument, 
\begin{eqnarray}
\text{i.e.: } &\mathcal{S}(d) \propto& \left(\frac{N_{\text{ON}}(d)}{\sqrt{N_{\text{OFF}}(d)}}\right)^{-1}.
\end{eqnarray}
Based on \autoref{eq:differentialSensitivity}, \autoref{eq:relativeAcceptance} can be re-written as:
\begin{equation}\label{eq:AcceptancevsSen}
\epsilon_{\text{CTA}}\left(d\right) = \frac{1}{\delta^{2}\left(d\right)}.
\end{equation}
\autoref{fig:relativeAcceptance} shows $\epsilon_{\text{CTA}}$, for the lower energy range from the northern CTA array shown
in \href{https://www.cta-observatory.org/science/cta-performance/cta-performance-archive1/}{cta-observatory.org/science/cta-performance/cta-performance-archive1/} 
(labeled as \emph{CTA North 50-80 GeV})\footnote{
This choice is particularly interesting since, in the next section, we apply the method for indirect DM searches on dSphs where
this energy range is typically considered among the most relevant for several DM models and also, because dSphs are particularly well 
observed from the northern hemisphere.}.
We assume here that $\epsilon$ for the ``CTA North 50-80 GeV'' is valid for the full CTA array. 
In reality CTA will be formed by IACTs of different kinds, with a different $\epsilon$ for each telescope type,
where the method would still be valid to optimize the pointing of each telescope type individually.
\newline

\noindent
We also stress that, based on $\epsilon$ in~\autoref{fig:relativeAcceptance}, we cannot compare the absolute acceptances between 
MAGIC and CTA. 

\section{Pointing optimization}
\label{sec:JDOptimization}
\begin{figure}[t]
  \centering
  \begin{subfigure}[b]{0.47\textwidth}
 \includegraphics[width=1.\linewidth]{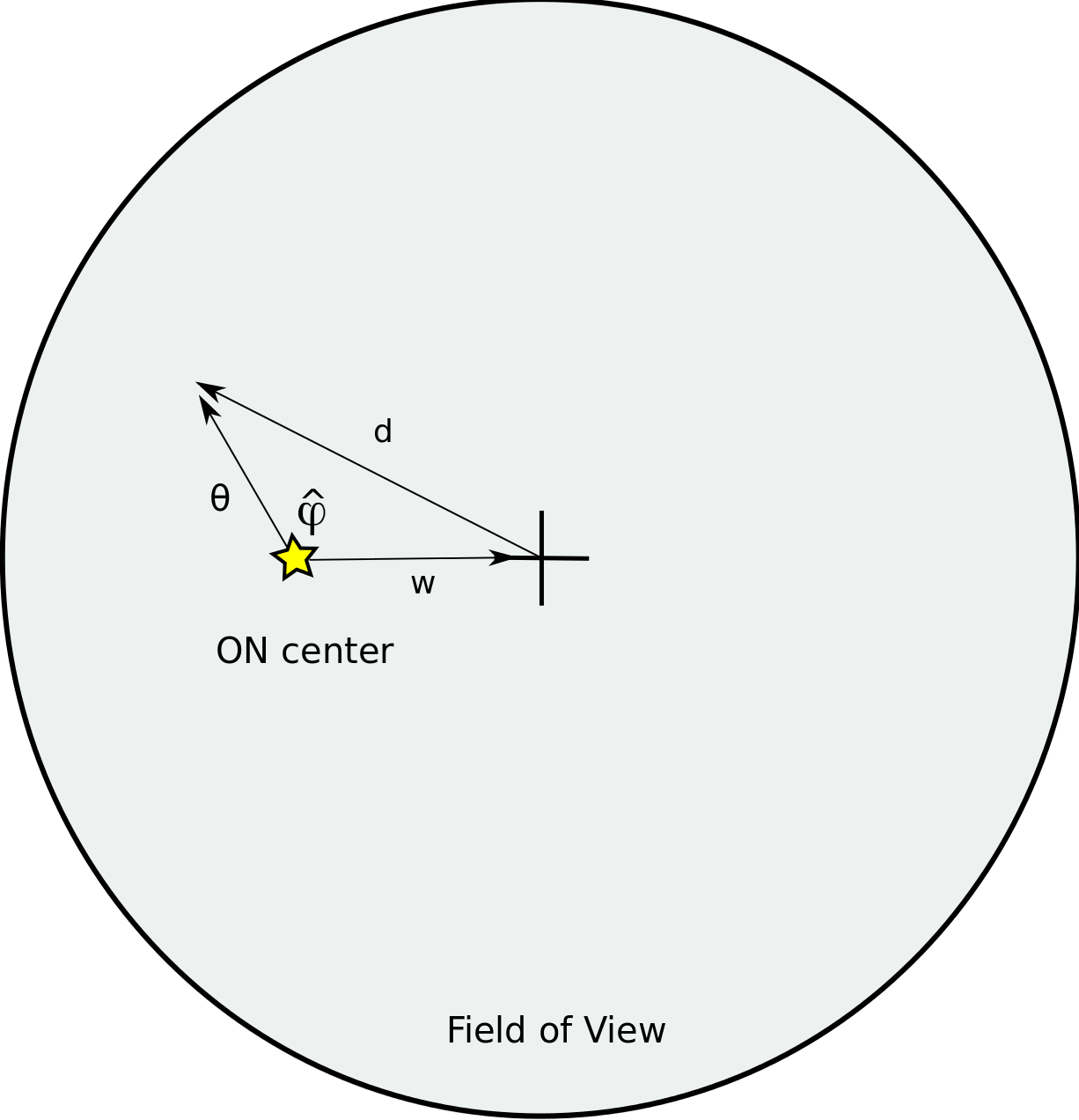}
    \caption{}
    \label{fig:wobbleModeVariables2_A}
  \end{subfigure}
  ~~
  \begin{subfigure}[b]{0.47\textwidth}
    \includegraphics[width=1.\linewidth]{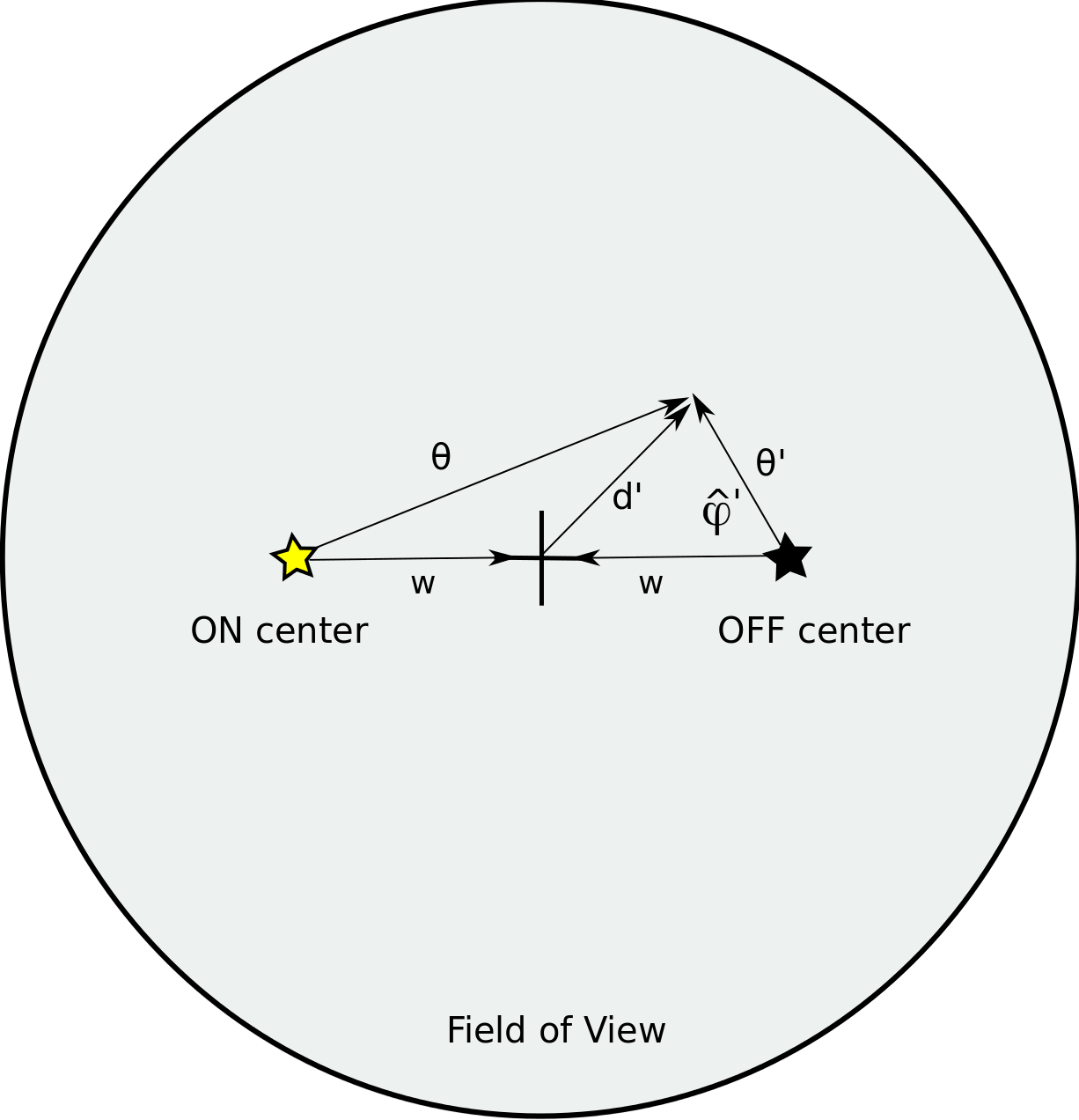}
    \caption{}
    \label{fig:wobbleModeVariables2_B}
  \end{subfigure}
\caption[Wobble mode variables]{\label{fig:wobbleModeVariables2} 
Definition of variables: (left) $w$ is the distance between the center of a source (yellow star) and the center of the FoV.
$\theta$ is the distance between the center of the source and any point of the FoV
(defined as the telescope nominal pointing direction). 
$d$ is the distance of any point in the FoV and the center of the FoV.
The three quantities are related by $\varphi$, the angle formed by the vectors $\vec{\theta}$ and $\vec{w}$.
(right) as for the case of (a) but $\theta'$, $\varphi'$ and $d'$ defined w.r.t. the center of the OFF region,
i.e. a direction mirroring the ON center w.r.t. the FoV center, i.e.
is located at the same $w$ but at the opposite side of the FoV (black star). 
Note that $d(w,\theta,\varphi)=d'(w,\theta',\varphi')$.}
\end{figure}
\noindent
We define the quality factor $\q$ ($Q$-factor) as the number of gamma rays from a given source 
in the ON region divided by the square-root of the number of background events within the same region.
As an illustration, in the following lines we compute partial values of $\q$, alternatively taking into account 
only one of the effects entering the global definition, shown at the end.
\newline

\noindent
Assuming the main contribution of background to be flat along the FoV, $\q$  can be written as:
\begin{eqnarray} \label{eq:QFactor}
 \nonumber
 \q \left(\thetac\right) &=& \frac{ \int_{\theta=0}^{\thetac}\int_{\varphi=0}^{2\pi} \theta d\theta d\varphi \hspace{0.1cm} \prof\left(\theta\right)}{\sqrt{\int_{\theta=0}^{\thetac}\int_{\varphi=0}^{2\pi} \theta d\theta d\varphi}} %\\
 = \frac{ \int_{\DomON} \domON \hspace{0.1cm} \prof\left(\theta\right)}{\sqrt{\int_{\DomON} \domON}} \; ,
\nonumber \\
\text{where } && \domON = \theta d\theta d\varphi,
\end{eqnarray}
$\theta$ and $\varphi$ are the circular coordinates w.r.t to the center of the ON region (see~\autoref{fig:wobbleModeVariables2_A}),
and the \emph{signal profile} $\prof$ is proportional to the number of gamma rays $N$ arriving from a given direction $d\Omega$ as:
\begin{equation}\label{eq:dNdOmegaProfile}
\prof=\mathcal{A}\cdot{dN}/{d\Omega}.
\end{equation} 
$\DomON$ is the region defined by: $\theta$ between $0$ and $\thetac$; and $\varphi$ between $0$ and $2\pi$.
$\q$ is maximal when the signal dominates the most over the background fluctuations
and we can therefore optimize the sensitivity of our observations by maximizing $\q$. 
Because we are interested only in maximizing $\q$ and not in its absolute value,
we fix the value of $\mathcal{A}$ such that $\q_{\text{max}}=1$.
\newline

\noindent
In general, given a signal profile $\prof$, $\q$ increases with $\thetac$ up to a point where
mostly background events start to be integrated, and $\q$ decreases.
We define $\theta_{\text{opt}}$ as the value of $\thetac$ that maximizes $\q$ ($\q\left(\theta_{\text{opt}}\right)=\q_{\text{max}}$). 
We also compute an interval around $\theta_{\text{opt}}$ for which $\q$ is within 30\% of the maximum \citep[which corresponds to 
the assumed systematic uncertainty in the determination of absolute fluxes with MAGIC, see][]{Aleksic:2014lkm}.
\subsection{$\qac$: Finite Acceptance}
%\subsection{Finite Acceptance}
\noindent
As introduced in~\autoref{sec:JDInstrument}, 
the off-axis performance of IACTs degrades towards the edges of the FoV. 
For wobble mode observations, it is important to take into account $\epsilon$ in order to determine the optimal $w$ and $\theta_{\text{c}}$.
We define $\qac$ as;
\begin{eqnarray}\label{eq:QFactor_acceptance}
    \qac\left(w,\thetac\right) 
    &=&  \frac{\int_{\DomON} \; \domON \; \prof\left(\theta\right)
    \epsilon\left(d\right)}{\sqrt{\int_{\DomON} \; \domON \; \epsilon\left(d\right)}}; \\
\nonumber \\%   \text{where }  d\Omega'&=& \theta' d\theta' d\varphi' \\
\nonumber \text{where }  \\
\nonumber d &=& \sqrt{\theta^{2}+w^{2}-2 \cdot \theta \cdot w \cdot \cos(\varphi)} .
\end{eqnarray}
For large values of $w$ and/or $\thetac$, $\epsilon$ is low, and hence $\qac$ decreases.
\subsection{$\qle$: Leakage Effect}
\noindent
Another effect to consider is that
for low values of $w$, ON and OFF regions are close to each other, and depending on $\prof$, it may not be possible to 
define a \emph{signal-free} OFF region (i.e. signal events ``leak'' into the background region).
This \emph{leakage effect} is exemplified in~\autoref{fig:wobbleModeVariables}, where gamma-ray events from 
$\prof$ (green circular area aligned with ON) are expected to be reconstructed inside OFF.
In order to take this effect into account we define $\qle$; 
\begin{eqnarray}\label{eq:QFactor_leakage}
\nonumber
\qle\left(w,\thetac\right) 
    &=&  \frac{\int_{\DomON}  \domON \; \prof\left(\theta\right)
    -\int_{\DomOFF}  \domOFF \; \prof\left(\theta\right)}{\sqrt{\int_{\DomON} \domON}}\\
\text{where }&& \domOFF= \theta' d\theta' d\varphi', \\
\nonumber
&& \theta= \sqrt{(2w)^{2} + \theta'^{2} + 2\cdot (2w) \cdot \theta' \cdot \cos{\varphi'}},
\end{eqnarray}
\noindent
$\theta'$ and $\varphi'$ are the polar coordinates w.r.t. the OFF center,
and $\DomOFF$ is the region defined by: $\theta$' between $0$ and $\thetac$; and $\varphi$' between $0$ and $2\pi$ (see~\autoref{fig:wobbleModeVariables2_B}).
Note that even while integrating over $\DomOFF$, $\prof$ has to be evaluated w.r.t. the ON (and source) center
(yellow star and $\theta$ in~\autoref{fig:wobbleModeVariables2_B}).
\newline

\noindent
Large values of $w$ are favoured since the distance between ON and OFF regions gets larger with $w$
(and the leakage between both regions smaller).
\paragraph{Alternative}A more correct definition of $\qle$ would be:
\begin{eqnarray}\label{eq:QFactor_leakage_B}
\qle\left(w,\thetac\right) 
    &=&  \frac{\int_{\DomON}  \domON \; \prof\left(\theta\right)}{\sqrt{\int_{\DomOFF}\domOFF \left(\mathcal{B} + \prof\left(\theta\right)\right)}} %\hspace{1cm} \text{(NOT)}
\end{eqnarray}
\noindent
for which we would need to know the relative intensities of signal and background components (parametrized in 
\autoref{eq:QFactor_leakage_B} by $\mathcal{B}$).
Generally the intensity of the signal is unknown and this definition is of no practical use. % for our purposes.
\subsection{$\qpsf$: Point Spread Function}
\noindent
Finally, we also take into account the finite angular resolution of IACTs.
We treat this effect convolving  $\prof$ with the point spread function (PSF) of the instrument,
approximated here by a circular-symmetric two-dimensional Gaussian ($\mathcal{G}_{\text{2D}}$)\footnote{
Note that the PSF of the instruments %presented in here 
is assumed to be independent of $d$.}:
\begin{eqnarray}
 \label{eq:psfConvolution}
\nonumber
 \prof' \left( \theta \right) &=& \int_{\Dom''}  \dom'' \; \prof\left(\theta'',\varphi''\right) \; \mathcal{G}_{\text{2D}}\left(\theta, \varphi, \theta'', \varphi'', \sigma\right)\\
\nonumber
 &=& \int_{\theta''=0}^{\infty} \theta'' d\theta'' \int_{\varphi''=0}^{2\pi} d\varphi'' \; \prof\left(\theta'',\varphi''\right) \; \frac{1}{2\pi\sigma^{2}}e^{-\frac{1}{2}\left[\frac{\left(\theta_{x}-\theta_{x}''\right)^{2}+\left(\theta_{y}-\theta_{y}''\right)^{2}}{\sigma^{2}}\right]}
\end{eqnarray}
where $\prof'$ is the differential gamma-ray rate smeared with the instrument PSF,
$\theta''$ and $\varphi''$ are the coordinates w.r.t. the center of the source,
$\sigma$ is the standard deviation,
and in the integral, $\theta$ and $\varphi$ have been expressed in Cartesian coordinates as,
\begin{eqnarray}
\nonumber
  \theta_{x} = \theta \cos \; \left(\varphi\right), & &  \theta_{y} = \theta \sin \; \left(\varphi\right);\\
  \theta''_{x} = \theta'' \cos \; \left(\varphi''\right), & &  \theta''_{y} = \theta'' \sin \; \left(\varphi''\right).
 %\text{where, }&& \dom''= \theta'' d\theta'' d\varphi'',
\end{eqnarray}
We define then $\qpsf$ as; 
\begin{eqnarray}\label{eq:QFactor_psf}
\qpsf\left(w,\thetac\right) 
    &=&  \frac{\int_{\DomON}  \domON \; \prof'\left(\theta,\varphi\right) }{\sqrt{\int_{\DomON} \domON}}.
\end{eqnarray} 
The effect of the PSF is dominant for point-like sources (smaller than the instrument PSF) however, it may 
also have a small impact on moderately extended sources.
\newline

\noindent
For the case considered in here, we set $\sigma$ (in~\autoref{eq:psfConvolution}) to $0.09^\circ$ for 
MAGIC~\citep[Figure 14 left in ][evaluated at $100$~GeV]{Aleksic:2014lkm}, 
and to $0.07^{\circ}$ for 
CTA~\citep[Figure 5 in][evaluated at $100$~GeV and using the relation $\sigma_{68}\approx 1.5\sigma$,
where $\sigma_{68}$ is the radius containing the $68$\% gamma-ray candidates from a point-like source w.r.t. to its center]{Hassan:2017paq}.
\subsection{$\qall$: ``Acceptance + Leakage + PSF'' Effect}
\begin{figure}[t]
\includegraphics[scale=0.4]{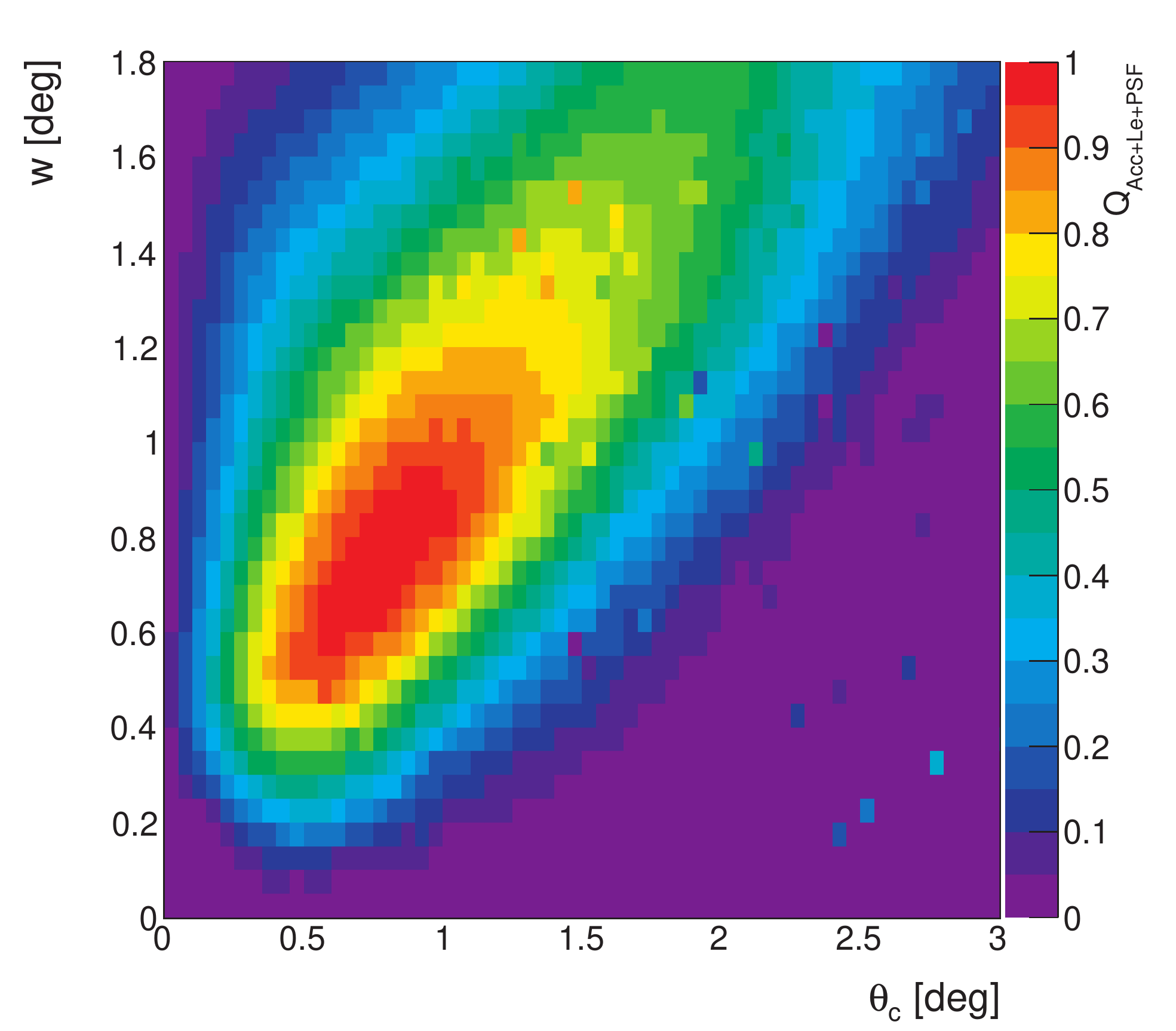}
\centering
\caption[$\qall$ vs $\thetac$ and $w$]{\label{fig:QFactor_Final}
$\qall$ as a function of $\thetac$ and $w$, computed as an example 
from ${dN}/{d\Omega}$ from the Coma dSph defined in~\cite{Bonnivard:2015xpq} and $\epsilon_{\text{MAGIC}}$.}
\end{figure}
\noindent
In general, we want to compute the optimal pointing strategy taking all effects into account, the finite acceptance of the instrument, the leakage of signal between ON and OFF,
and the finite angular resolution of the instrument.
For that, we define $\qall$
\begin{eqnarray}\label{eq:QFactor_Final}
\nonumber
\qall\left(w,\thetac\right)  &=& \q_{Ac + Le + _{PSF}} \left(w,\thetac\right)  \\
\nonumber \\
&=&\frac{\int_{\DomON}  \domON \; \prof'\left(\theta\right) \epsilon\left(d\right)
    -\int_{\DomOFF}  \domOFF \; \prof'\left(\theta\right) \epsilon\left(d'\right)}
    {\sqrt{\int_{\DomON}  \domON \; \epsilon\left(d\right)}}; \\
\nonumber \\
\nonumber \text{where } && d' = \sqrt{\theta'^{2}+w^{2}-2 \cdot \theta' \cdot w \cdot \cos(\varphi')} \; \; \; \; (=d) 
\end{eqnarray}
\autoref{fig:QFactor_Final} shows $\qall$ as a function of the observational variables
$w$ and $\thetac$ for the Coma dSph~\citep{Bonnivard:2015xpq}.
$\qall$ is a function of $\theta_{\text{c}}$ and $w$, and we define $w_{\text{opt}}$, $\theta_{\text{opt}}$ and their contour regions
$\Delta w_{\text{opt}}$ and $\Delta \theta_{\text{opt}}$, as 
the values that maximize $\qac$ and that $\qac$ is within 30\% of the maximum. 
% (in analogy to what we did in~\autoref{eq:QFactor}).
The acceptance and the leakage effect have opposed tendencies w.r.t. $\theta_{\text{opt}}$ and $w_{\text{opt}}$, and the optimal 
region defined $\qall$ defines a narrow region around them.

\section{Optimized pointing strategy for indirect Dark Matter searches}
\label{sec:JDDarkMatter}
\noindent
IACTs core science is focused on the study of the cosmic ray origin in
either Galactic or extragalactic targets, but it is well-known that cosmic
gamma rays constitute also a probe for several fundamental 
physics investigations~\citep[including DM searches, see e.g.][]{Doro:2012xx}.
We can use \autoref{eq:QFactor_Final} to optimize the %observations and analysis of the indirect DM~\citep{Bertone:2004pz}
search of Weakly-Interacting Massive Particles~~\citep[WIMPs, generic massive particles postulated to solve the DM problem, 
see][]{Boehm:2003hm,Griest:1989wd} with Cherenkov telescopes.
%WIMPs are generic massive particles postulated to solve the DM problem~\citep{Boehm:2003hm,Griest:1989wd}.
\newline

\noindent
The gamma-ray flux from annihilating (or decaying) WIMPs 
arriving at Earth from a given region of the sky ($\Dom$) can be factorized as
\begin{equation}\label{eq:flux}
\frac{d\Phi(E,\Delta\Omega)}{dE}=\ppterm \cdot J(\Delta\Omega), 
\end{equation}
\noindent
where $\sppterm$ is called
the \emph{particle-physics} factor, and depends on the nature of DM,
and $ J(\dom)$ is called the \emph{astrophysics} factor (or simply
$J$-factor), and depends on the target distance and the DM
distribution therein. These two factors read:
\begin{eqnarray}\label{eq:PPandJfactor}
\nonumber 
\ppterm & = & \frac{1}{4\pi}\frac{\aaa}{k\, \mdm^k}\frac{dN}{dE} (E) \\
\nonumber
J(\Dom) & = &  \int_{\dom}  d\Omega \ \frac{dJ(l,\Omega)}{d\Omega} \; ,\\
\nonumber \\
&\text{where }& \: \frac{dJ(l,\Omega)}{d\Omega} = \int_\mathrm{l.o.s.} dl\,  \rho^k(l,\Omega)
\end{eqnarray}
respectively, with
\begin{eqnarray}
\aaa = \sv, k=2 &, & \mbox{ for annihilating DM,} \nonumber\\
\aaa = \tdm^{-1}, k=1 &, & \mbox{ for decaying DM;} \nonumber 
\end{eqnarray}
$\sv$, $\tdm$ and $\mdm$ are the DM particle
velocity-averaged annihilation cross section, lifetime, and mass,
respectively; $dN/dE$ is the average gamma-ray spectrum of a DM
annihilation or decay event;
%, obtained from references \cite{Cirelli2011} and \cite{Cirelli2013}; 
and $\rho$ the DM density at a given sky direction $\Omega$ and distance
from Earth $l$. The integrals in the astrophysical factor run over the
region $\Dom$ and the line of sight, respectively;
$\rho$ is typically assumed to be spherically symmetric, i.e. $\rho=\rho(r)$. 
%For the rest of this work we will use the acronym $J_{\text{ann}}$ ($J_{\text{dec}}$)
%to refer to the $J$-factor for the annihilation (decay) case.
%\section{Results}
%\label{sec:Results}
%\input{004_Results}
\newline

\noindent
The DM signal profile $\prof$ is determined by the $J$-factor: 
\begin{equation}\label{eq:J2dNdOmega}
    %\frac{dN}{d\Omega}\propto\frac{dJ}{d\Omega}.
    \prof\propto\frac{dJ}{d\Omega}.
\end{equation}
The proportionality constant is absorbed in $\mathcal{A}$ (see~\autoref{eq:dNdOmegaProfile}).
Thus, using~\autoref{eq:J2dNdOmega} in~\autoref{eq:QFactor_Final}, we can optimize $w$ and $\thetac$ for DM observations.
\cite{Bonnivard:2015xpq} and \cite{Geringer-Sameth:2014yza} provide the $J$-factor for two sets 
of DM halos hosting Milky Way dSphs, 
as a function of the signal integration angle ($\theta_{\text{c}}$).
We apply the method to optimize the pointing strategy %of MAGIC and CTA
%for indirect DM searches.
%\newline
%
%\noindent
%Based on \cite{Bonnivard:2015xpq} and \cite{Geringer-Sameth:2014yza}, we numerically compute $dJ/d\Omega$ (hence $\prof$)
of annihilating and decaying DM for all available dSphs from both authors to be observed with MAGIC and CTA 
%the two possible type of candidates in~\autoref{eq:PPandJfactor} (
%;
%and based on $\overline{\mathscr{Q}}$, 
taking into account the three effects introduced 
in~\autoref{eq:QFactor_Final} (acceptance, leakage and PSF). 
\begin{table*}[t]
  \tiny
  \centering
  \begin{tabular}{|l|cc|cc||cc|cc|}
  \hline
   & \multicolumn{4}{c||}{MAGIC} & \multicolumn{4}{c|}{CTA} \\
  source & \multicolumn{2}{c}{$\theta_{\text{opt}}$} & \multicolumn{2}{c||}{$w_{\text{opt}}$} & \multicolumn{2}{c}{$\theta_{\text{opt}}$} & \multicolumn{2}{c|}{$w_{\text{opt}}$} \\
  \hline
  boo1   &0.20& (0.10, 0.45)	&0.30& (0.15, 0.60)	&0.20& (0.10, 0.65)	&0.55& (0.15, 1.16)\\
  car    &0.15& (0.10, 0.30)	&0.25& (0.10, 0.55)	&0.15& (0.10, 0.35)	&0.40& (0.10, 1.16)\\
  coma   &0.25& (0.10, 0.60)	&0.35& (0.20, 0.65)	&0.35& (0.15, 0.95)	&0.65& (0.30, 1.21)\\
  cvn1   &0.15& (0.10, 0.30)	&0.25& (0.10, 0.55)	&0.15& (0.10, 0.30)	&0.40& (0.10, 1.16)\\
  cvn2   &0.15& (0.10, 0.35)	&0.30& (0.10, 0.55)	&0.15& (0.10, 0.35)	&0.40& (0.10, 1.16)\\
  dra    &0.25& (0.10, 0.50)	&0.35& (0.20, 0.65)	&0.30& (0.10, 0.80)	&0.60& (0.25, 1.16)\\
  for    &0.15& (0.10, 0.30)	&0.25& (0.10, 0.55)	&0.15& (0.10, 0.35)	&0.45& (0.10, 1.16)\\
  her    &0.20& (0.10, 0.40)	&0.30& (0.15, 0.55)	&0.20& (0.10, 0.50)	&0.50& (0.15, 1.16)\\
  leo1   &0.15& (0.10, 0.30)	&0.25& (0.10, 0.55)	&0.15& (0.10, 0.30)	&0.35& (0.10, 1.16)\\
  leo2   &0.15& (0.10, 0.30)	&0.25& (0.10, 0.55)	&0.15& (0.10, 0.30)	&0.35& (0.10, 1.16)\\
  leo4   &0.15& (0.10, 0.30)	&0.25& (0.10, 0.55)	&0.15& (0.10, 0.35)	&0.45& (0.10, 1.16)\\
  leo5   &0.15& (0.10, 0.35)	&0.30& (0.10, 0.55)	&0.15& (0.10, 0.40)	&0.45& (0.10, 1.16)\\
  leot   &0.15& (0.10, 0.30)	&0.25& (0.10, 0.55)	&0.15& (0.10, 0.30)	&0.30& (0.10, 1.16)\\
  scl    &0.15& (0.10, 0.30)	&0.25& (0.10, 0.55)	&0.15& (0.10, 0.30)	&0.35& (0.10, 1.16)\\
  seg1   &0.25& (0.10, 0.50)	&0.35& (0.20, 0.60)	&0.30& (0.10, 0.75)	&0.55& (0.25, 1.16)\\  
  seg2   &0.25& (0.10, 0.55)	&0.35& (0.20, 0.65)	&0.35& (0.15, 0.95)	&0.65& (0.30, 1.21)\\
  sex    &0.25& (0.10, 0.55)	&0.35& (0.20, 0.65)	&0.35& (0.15, 0.85)	&0.60& (0.25, 1.21)\\
  uma1   &0.15& (0.10, 0.30)	&0.25& (0.10, 0.55)	&0.15& (0.10, 0.35)	&0.45& (0.10, 1.16)\\
  uma2   &0.35& (0.15, 0.75)	&0.45& (0.25, 0.75)	&0.50& (0.20, 1.10)	&0.70& (0.35, 1.26)\\
  umi    &0.15& (0.10, 0.30)	&0.25& (0.10, 0.55)	&0.15& (0.10, 0.30)	&0.40& (0.10, 1.16)\\
  wil1   &0.15& (0.10, 0.30)	&0.25& (0.10, 0.55)	&0.15& (0.10, 0.30)	&0.40& (0.10, 1.16)\\
  \hline
  \end{tabular}
  \caption[Optimal $\theta_{\text{c}}$ and $w$ for annihilation in Bonnivard]{\label{tab:optimalValues_Bonnivard_Annihilation}
  List of optimal pointing wobble distance ($w_\text{opt}$) and signal region radius ($\theta_{\text{opt}}$),
  and their contour regions defined within 30\% of the maximum of $\qall$,
  for annihilating WIMP based on ${dJ}/{d\Omega}$ taken from~\cite{Bonnivard:2015xpq}.
  First column show the dSph name (taken from~\cite{Bonnivard:2015xpq}).
  Second and third (fourth and fifth) show the optimal $\theta_{\text{c}}$ and $w$ for 
  observations with MAGIC (CTA).}
\end{table*}
\begin{table*}[t]
  \tiny
  \centering
  \begin{tabular}{|l|cc|cc||cc|cc|}
  \hline
   & \multicolumn{4}{c||}{MAGIC} & \multicolumn{4}{c|}{CTA} \\
  source & \multicolumn{2}{c}{$\theta_{\text{opt}}$} & \multicolumn{2}{c||}{$w_{\text{opt}}$} & \multicolumn{2}{c}{$\theta_{\text{opt}}$} & \multicolumn{2}{c|}{$w_{\text{opt}}$} \\
  \hline
  boo&	0.20& (0.10, 0.40)	&0.30& (0.15, 0.55)	&0.25& (0.10, 0.45)	&0.35& (0.20, 1.16)\\
  car&	0.15& (0.10, 0.35)	&0.30& (0.10, 0.55)	&0.15& (0.10, 0.45)	&0.45& (0.15, 1.16)\\
  coma&	0.15& (0.10, 0.30)	&0.25& (0.10, 0.55)	&0.15& (0.10, 0.30)	&0.30& (0.10, 1.16)\\
  cvn1&	0.15& (0.10, 0.40)	&0.30& (0.15, 0.55)	&0.20& (0.10, 0.45)	&0.35& (0.15, 1.16)\\
  cvn2&	0.15& (0.05, 0.25)	&0.20& (0.10, 0.55)	&0.15& (0.05, 0.20)	&0.30& (0.10, 1.16)\\
  dra&	0.25& (0.10, 0.55)	&0.35& (0.20, 0.65)	&0.35& (0.15, 0.85)	&0.60& (0.25, 1.16)\\
  for&	0.15& (0.10, 0.35)	&0.25& (0.10, 0.55)	&0.15& (0.10, 0.35)	&0.35& (0.10, 1.16)\\
  her&	0.15& (0.05, 0.25)	&0.25& (0.10, 0.55)	&0.15& (0.05, 0.25)	&0.30& (0.10, 1.16)\\
  leo1&	0.15& (0.10, 0.35)	&0.25& (0.10, 0.55)	&0.15& (0.10, 0.40)	&0.35& (0.15, 1.16)\\
  leo2&	0.15& (0.05, 0.25)	&0.25& (0.10, 0.55)	&0.15& (0.05, 0.25)	&0.30& (0.10, 1.16)\\
  leo4&	0.15& (0.05, 0.25)	&0.20& (0.10, 0.55)	&0.15& (0.05, 0.20)	&0.30& (0.10, 1.16)\\
  leo5&	0.15& (0.05, 0.20)	&0.20& (0.10, 0.55)	&0.10& (0.05, 0.20)	&0.35& (0.10, 1.16)\\
  leot&	0.15& (0.05, 0.20)	&0.20& (0.10, 0.55)	&0.10& (0.05, 0.20)	&0.35& (0.10, 1.16)\\
  scl&	0.15& (0.10, 0.35)	&0.30& (0.15, 0.55)	&0.15& (0.10, 0.45)	&0.40& (0.15, 1.16)\\
  seg1&	0.15& (0.10, 0.30)	&0.25& (0.10, 0.55)	&0.15& (0.10, 0.35)	&0.30& (0.10, 1.16)\\
  seg2&	0.15& (0.05, 0.25)	&0.25& (0.10, 0.55)	&0.15& (0.05, 0.25)	&0.30& (0.10, 1.16)\\
  sex&	0.30& (0.10, 0.65)	&0.40& (0.20, 0.70)	&0.50& (0.15, 1.10)	&0.75& (0.35, 1.21)\\
  uma1&	0.15& (0.10, 0.35)	&0.25& (0.15, 0.55)	&0.20& (0.10, 0.40)	&0.35& (0.15, 1.16)\\
  uma2&	0.25& (0.10, 0.45)	&0.30& (0.20, 0.55)	&0.30& (0.10, 0.50)	&0.40& (0.20, 1.16)\\
  umi&	0.15& (0.05, 0.25)	&0.25& (0.10, 0.55)	&0.15& (0.05, 0.25)	&0.35& (0.10, 1.16)\\
  \hline
  \end{tabular}
  \caption[Optimal $\theta_{\text{c}}$ and $w$ for annihilation in Geringer]{\label{tab:optimalValues_Geringer_Annihilation}
  List of optimal pointing wobble distance ($w_{\text{opt}}$) and signal region radius ($\theta_{\text{opt}}$)
  for annihilating WIMP based on ${dJ}/{d\Omega}$~\cite{Geringer-Sameth:2014yza}.
  Column description can be found in~\autoref{tab:optimalValues_Bonnivard_Annihilation}. }
\end{table*}
\newline

\noindent
%Figure \ref{fig:JFactor_Annihilation} shows the J-factors for annihilation ($J_{ann}$) as a function of $\thetac$ 
%for all available dSphs in \citep[left]{Bonnivard:2015xpq} and \citep[right]{Geringer-Sameth:2014yza}. 
We focus first on the \emph{annihilation} case, and provide 
the optimal values $w_{\text{opt}}$, $\theta_{\text{opt}}$, and their 30\% variation ranges 
in~\autoref{tab:optimalValues_Bonnivard_Annihilation} and \ref{tab:optimalValues_Geringer_Annihilation}.
\newline

\noindent 
For the case of MAGIC, we note how $w_{\text{opt}}$ is systematically lower than $w_{\text{MAGIC}}=0.4^{\circ}$.
This is the case for most point-like sources,
for which, in the case of the standard analysis of MAGIC, 3 different off regions are considered,
and therefore, for the same $w$, the distance between these OFF regions and the ON is smaller.
There are a few cases in which the source appears to be moderately extended for 
MAGIC, i.e.~\emph{uma2}~\citep[in ][]{Bonnivard:2015xpq}
or \emph{sex}~\citep[in ][]{Geringer-Sameth:2014yza}.
%We also note the disagreement between these and the optimal values obtained in here for the source considered in both publications.
The discrepancies between the optimal values obtained (for the same source) from the two authors show the large uncertainties 
affecting the DM profiles.
%however, this matter is out of the scope of this work.
Finally, it should also be said that, for the sake of simplicity, the method does not take into account systematic effects that may affect the real analysis.
For instance, the systematic error on the background estimation, is proportional  
to the number of OFF events.
This means that for two different configurations (two different $w$ and $\theta_{\text{c}}$ pairs) with similar $\q$, 
we should give priority to the one with lower $\theta_{\text{c}}$ (lower statistics).
%Once systematic effects taken into account, the sensitivity is going to be better.
\newline

\noindent
For the case of CTA, our results can be taken as reference to schedule future observations.
However two caveats should be considered:
1)~CTA will be composed of two sites, one operating in the North (\emph{CTAN}) hemisphere and one in the south
(\emph{CTAS}) however, we treated all dSphs with the same instrument acceptance regardless of their position in the sky;  
2)~Each CTA site (CTAN and CTAS) will be integrated by, up to, three different types of telescope and hence, once CTA
analysis scheme is defined, a proper optimization could be performed for the pointing of each telescope using our code.
\newline
\begin{table*}[t]
  \tiny
  \centering
  \begin{tabular}{|l|cc|cc||cc|cc|}
  \hline
   & \multicolumn{4}{c||}{MAGIC} & \multicolumn{4}{c|}{CTA} \\
  source & \multicolumn{2}{c}{$\theta_{\text{opt}}$} & \multicolumn{2}{c||}{$w_{\text{opt}}$} & \multicolumn{2}{c}{$\theta_{\text{opt}}$} & \multicolumn{2}{c|}{$w_{\text{opt}}$} \\
  \hline
  boo1 	& 0.50 & (0.20, 0.85) 	& 0.50 & (0.30, 0.90)	& 0.65 & (0.25, 1.30) 	& 0.85 & (0.45, 1.30)\\
  car 	& 0.30 & (0.10, 0.65) 	& 0.40 & (0.20, 0.70)	& 0.50 & (0.15, 1.10) 	& 0.75 & (0.35, 1.30)\\
  coma 	& 0.50 & (0.25, 0.85) 	& 0.50 & (0.35, 0.90)	& 0.70 & (0.30, 1.70) 	& 0.91 & (0.50, 1.40)\\
  cvn1 	& 0.25 & (0.10, 0.65) 	& 0.40 & (0.20, 0.70)	& 0.45 & (0.15, 1.10) 	& 0.70 & (0.35, 1.20)\\
  cvn2 	& 0.30 & (0.15, 0.65) 	& 0.40 & (0.20, 0.70)	& 0.45 & (0.15, 1.00) 	& 0.70 & (0.30, 1.20)\\
  dra 	& 0.50 & (0.20, 0.95) 	& 0.55 & (0.35, 0.90)	& 0.75 & (0.25, 1.20) 	& 0.75 & (0.50, 1.30)\\
  for 	& 0.35 & (0.15, 0.70) 	& 0.45 & (0.25, 0.75)	& 0.50 & (0.20, 1.10) 	& 0.70 & (0.35, 1.30)\\
  her 	& 0.50 & (0.20, 0.95) 	& 0.55 & (0.30, 0.90)	& 0.65 & (0.25, 1.20) 	& 0.80 & (0.45, 1.30)\\
  leo1 	& 0.25 & (0.10, 0.55) 	& 0.35 & (0.20, 0.65)	& 0.35 & (0.15, 1.00) 	& 0.70 & (0.30, 1.20)\\
  leo2 	& 0.20 & (0.10, 0.50) 	& 0.30 & (0.15, 0.60)	& 0.25 & (0.10, 0.75) 	& 0.60 & (0.20, 1.20)\\
  leo4 	& 0.30 & (0.10, 0.65) 	& 0.40 & (0.20, 0.65)	& 0.50 & (0.15, 1.00) 	& 0.70 & (0.35, 1.20)\\
  leo5 	& 0.30 & (0.10, 0.65) 	& 0.40 & (0.20, 0.70)	& 0.50 & (0.15, 1.10) 	& 0.70 & (0.35, 1.20)\\
  leot 	& 0.20 & (0.10, 0.40) 	& 0.30 & (0.15, 0.60)	& 0.20 & (0.10, 0.50) 	& 0.50 & (0.15, 1.20)\\
  scl 	& 0.25 & (0.10, 0.50) 	& 0.35 & (0.20, 0.65)	& 0.30 & (0.10, 0.90) 	& 0.65 & (0.25, 1.20)\\
  seg1 	& 0.50 & (0.15, 0.95) 	& 0.55 & (0.30, 0.90)	& 0.65 & (0.25, 1.30) 	& 0.85 & (0.45, 1.30)\\
  seg2 	& 0.70 & (0.30, 1.10) 	& 0.65 & (0.40, 1.10)	& 0.70 & (0.30, 1.30) 	& 0.85 & (0.45, 1.40)\\
  sex 	& 0.65 & (0.30, 1.20) 	& 0.70 & (0.40, 1.00)	& 0.70 & (0.30, 5.50) 	& 0.85 & (0.50, 1.40)\\
  uma1 	& 0.30 & (0.15, 0.65) 	& 0.40 & (0.20, 0.70)	& 0.50 & (0.20, 1.10) 	& 0.70 & (0.35, 1.30)\\
  uma2 	& 5.60 & (1.10, 5.60) 	& 1.20 & (1.20, 1.20)	& 6.50 & (0.60, 6.50) 	& 1.40 & (1.40, 1.40)\\
  umi 	& 0.30 & (0.10, 0.65) 	& 0.40 & (0.20, 0.70)	& 0.50 & (0.15, 1.10) 	& 0.70 & (0.35, 1.30)\\
  wil1 	& 0.30 & (0.10, 0.70) 	& 0.45 & (0.20, 0.70)	& 0.50 & (0.15, 1.10) 	& 0.70 & (0.35, 1.20)\\
  \hline
  \end{tabular}
  \caption[Optimal $\theta_{\text{c}}$ and $w$ for decay in Bonnivard]{\label{tab:optimalValues_Bonnivard_Decay}
  List of optimal pointing wobble distance ($w_{\text{opt}}$) and signal region radius ($\theta_{\text{opt}}$)
  for decaying WIMP based on ${dJ}/{d\Omega}$ based on~\cite{Bonnivard:2015xpq}.
  Column description can be found in~\autoref{tab:optimalValues_Bonnivard_Annihilation}. }
\end{table*}
\begin{table*}[t]
  \tiny
  \centering
  \begin{tabular}{|l|cc|cc||cc|cc|}
  \hline
   & \multicolumn{4}{c||}{MAGIC} & \multicolumn{4}{c|}{CTA} \\
  source & \multicolumn{2}{c}{$\theta_{\text{opt}}$} & \multicolumn{2}{c||}{$w_{\text{opt}}$} & \multicolumn{2}{c}{$\theta_{\text{opt}}$} & \multicolumn{2}{c|}{$w_{\text{opt}}$} \\
  \hline
  boo	&0.25 & (0.15, 0.45)	&0.30 & (0.20, 0.55)	&0.30 & (0.15, 0.50)	&0.40 & (0.20, 1.16)\\
  car	&0.30 & (0.15, 0.65)	&0.40 & (0.25, 0.70)	&0.45 & (0.15, 0.95)	&0.65 & (0.35, 1.16)\\
  coma	&0.20 & (0.10, 0.35)	&0.25 & (0.15, 0.55)	&0.15 & (0.10, 0.35)	&0.30 & (0.10, 1.16)\\
  cvn1	&0.25 & (0.10, 0.45)	&0.30 & (0.20, 0.60)	&0.30 & (0.15, 0.50)	&0.40 & (0.20, 1.16)\\
  cvn2	&0.15 & (0.05, 0.25)	&0.25 & (0.10, 0.55)	&0.15 & (0.05, 0.20)	&0.30 & (0.10, 1.16)\\
  dra	&0.80 & (0.35, 1.15)	&0.75 & (0.45, 1.00)	&0.65 & (0.30, 1.15)	&0.75 & (0.45, 1.21)\\
  for	&0.25 & (0.10, 0.50)	&0.35 & (0.20, 0.60)	&0.30 & (0.15, 0.65)	&0.50 & (0.20, 1.16)\\
  her	&0.15 & (0.10, 0.30)	&0.25 & (0.10, 0.55)	&0.15 & (0.10, 0.30)	&0.30 & (0.10, 1.16)\\
  leo1	&0.25 & (0.10, 0.45)	&0.30 & (0.20, 0.55)	&0.25 & (0.10, 0.45)	&0.35 & (0.20, 1.16)\\
  leo2	&0.15 & (0.05, 0.25)	&0.25 & (0.10, 0.55)	&0.15 & (0.05, 0.25)	&0.30 & (0.10, 1.16)\\
  leo4	&0.15 & (0.05, 0.25)	&0.25 & (0.10, 0.55)	&0.15 & (0.05, 0.25)	&0.30 & (0.10, 1.16)\\
  leo5	&0.15 & (0.05, 0.25)	&0.20 & (0.10, 0.55)	&0.10 & (0.05, 0.20)	&0.35 & (0.10, 1.16)\\
  leot	&0.15 & (0.05, 0.25)	&0.20 & (0.10, 0.55)	&0.15 & (0.05, 0.20)	&0.30 & (0.10, 1.16)\\
  scl	&0.30 & (0.15, 0.65)	&0.40 & (0.20, 0.65)	&0.40 & (0.15, 0.90)	&0.65 & (0.30, 1.21)\\
  seg1	&0.20 & (0.10, 0.35)	&0.25 & (0.15, 0.55)	&0.20 & (0.10, 0.35)	&0.30 & (0.15, 1.16)\\
  seg2	&0.15 & (0.05, 0.25)	&0.25 & (0.10, 0.55)	&0.15 & (0.05, 0.25)	&0.30 & (0.10, 1.16)\\
  sex	&0.90 & (0.45, 1.35)	&0.85 & (0.55, 1.20)	&0.80 & (0.35, 1.45)	&0.91 & (0.55, 1.31)\\
  uma1	&0.25 & (0.10, 0.45)	&0.30 & (0.20, 0.55)	&0.25 & (0.10, 0.45)	&0.35 & (0.20, 1.16)\\
  uma2	&0.30 & (0.15, 0.50)	&0.35 & (0.20, 0.60)	&0.35 & (0.15, 0.55)	&0.40 & (0.25, 1.16)\\
  umi	&0.20 & (0.10, 0.45)	&0.30 & (0.15, 0.60)	&0.25 & (0.10, 0.60)	&0.50 & (0.20, 1.16)\\
  \hline
  \end{tabular}
  \caption[Optimal $\theta_{\text{c}}$ and $w$ for decay in Geringer]{\label{tab:optimalValues_Geringer_Decay}
  List of optimal pointing wobble distance ($w_{\text{opt}}$) and signal region radius ($\theta_{\text{opt}}$)
  for decaying WIMP based on ${dJ}/{d\Omega}$ based on~\cite{Geringer-Sameth:2014yza}.
  Column description can be found in~\autoref{tab:optimalValues_Bonnivard_Annihilation}. }
\end{table*}

\noindent
Tables \ref{tab:optimalValues_Bonnivard_Decay} and \ref{tab:optimalValues_Geringer_Decay} show 
the optimal values ($w_{\text{opt}}$, $\theta_{\text{opt}}$) for the \emph{decay} case.
\newline

\noindent
Most of these sources are considered to be rather extended
(this is expected given the dependence on $\rho$ in~\autoref{eq:PPandJfactor}).

\section{Summary and Discussion}
\label{sec:Conclusions}
\noindent
In this work, we have proposed a method to optimize the pointing strategy and analysis for extended sources observed by IACTs.
The method provides the optimal offset and signal integration distances ($w_{\text{opt}}, \theta_{\text{opt}}$) taking into account: 
the off-axis performance and the angular resolution of the instrument,
and the profile of the source under observation.
The method has a potential use in scheduling new observations, but can also be used to optimize the analysis cut $\theta_{\text{c}}$
(typically used by the community as a cut on $\theta^{2}$)
for data already taken.
We focus on the case of indirect DM searches, and provide optimal pointing strategies for indirect DM searches 
on a set of dSph to be observed with MAGIC and CTA.
\newline

\noindent
We have implemented the method in a tool that is freely distributed, open source software, accessible from:
{\scriptsize
\begin{itemize}
    \item[]  \href{https://github.com/IndirectDarkMatterSearchesIFAE/}{\texttt{https://github.com/IndirectDarkMatterSearchesIFAE/}}
\end{itemize}
}
\noindent
A released version (\emph{V1.0}), with which the results shown in this paper were computed, can be accessed by:\\
{\scriptsize
\begin{itemize}
    \item[] \texttt{\$~git clone https://github.com/IndirectDarkMatterSearchesIFAE/ObservationOptimization.git}
    \item[] \texttt{\$~git checkout V1.0}
\end{itemize}
}
\noindent
The package is provided with tutorials in order to acquire the basic skills required
to reproduce the results shown here.
The software is flexible enough so that new sources (not necessary related to DM) or telescopes can be defined easily.
This provides an easy, fast, and powerful tool for planning new observations with IACTs.
\section*{Acknowledgement}
We thank M. Doro, for encouraging us from the very beginning to develop further this idea. 
We also thank T. Hassan, without whom, this work would not have been possible.

\bigskip
This research has made use of the CTA instrument response functions provided
by the CTA Consortium and Observatory, see \\
%\begin{itemize}
 %\item[] 
 \href{http://www.cta-observatory.org/science/cta-performance/}{http://www.cta-observatory.org/science/cta-perfomance/} 
%\end{itemize}
%(version \emph{prod3b-v1}) 
for more details.
This paper has gone through internal review by the CTA Consortium.
\bigskip
\bigskip
%
%\clearpage
\bibliography{999_Library}
%\clearpage
%\appendix 
%\section*{Previous Values}
%\input{888_appendix}
\end{document}